\documentclass[%
reprint,
longbibliography,
 preprintnumbers,
 nofootinbib,
 amsmath,amssymb,
 aps,
 prl,
]{revtex4-1}
\usepackage{psfrag,slashed,cancel,array,graphicx,todonotes,hyperref}
\usepackage{subfigure}
\usepackage[utf8]{inputenc}
\usepackage{mathtools}
\usepackage{mathrsfs}
\usepackage{multirow}
\usepackage{braket}
\usepackage{bbm}
\usepackage{physics}
\usepackage{mathtools}
 \usepackage{delimset} 
\usepackage{booktabs} 
\usepackage[normalem]{ulem}
\hypersetup{pdftitle={},pdfcreator={},linkcolor=[rgb]{0.15,0.35,0.75},colorlinks=true,citecolor=[rgb]{0.675,0,0.2},urlcolor=[rgb]{0.15,0.35,0.65}}

\let\oldPhi=\Phi
\let\oldPsi=\Psi
\let\oldGamma=\Gamma
\let\oldDelta=\Delta
\let\oldSigma=\Sigma
\let\oldTheta=\Theta
\let\oldPi=\Pi
\let\oldUpsilon=\Upsilon
\renewcommand{\Phi}{\mathnormal{\oldPhi}}
\renewcommand{\Psi}{\mathnormal{\oldPsi}}
\renewcommand{\Gamma}{\mathnormal{\oldGamma}}
\renewcommand{\Sigma}{\mathnormal{\oldSigma}}
\renewcommand{\Delta}{\mathnormal{\oldDelta}}
\renewcommand{\Theta}{\mathnormal{\oldTheta}}
\renewcommand{\Pi}{\mathnormal{\oldPi}}
\renewcommand{\Upsilon}{\mathnormal{\oldUpsilon}}

\newcommand{\superN}{\mathcal{N}}

\newcommand{\gen}[1]{\mathrm{#1}}
\newcommand{\levo}[1]{ \gen{\widehat #1}}

\newcommand{\Eval}{s} 
\makeatletter
\newlength{\apb@width}
\newcommand{\autoparbox}[2][c]{\settowidth{\apb@width}{#2}\parbox[#1]{\apb@width}{#2}}
\newcommand{\includegraphicsbox}[2][]{\autoparbox{\includegraphics[#1]{#2}}}
\makeatother

\ifx\genfrac\sdflkaj

\else

\fi
\newcommand{\sfrac}[2]{{\textstyle\frac{#1}{#2}}}

\makeatletter
\def\mr@ignsp#1 {\ifx\:#1\@empty\else #1\expandafter\mr@ignsp\fi}%
\newcommand{\multiref}[1]{\begingroup
\xdef\mr@no@sparg{\expandafter\mr@ignsp#1 \: }%
\def\mr@comma{}%
\@for\mr@refs:=\mr@no@sparg\do{\mr@comma\def\mr@comma{,}\ref{\mr@refs}}%
\endgroup}
\makeatother

\newcommand{\hypref}[2]{\ifx\href\asklfhas #2\else\href{#1}{#2}\fi}

\newcommand{\Figref}[1]{Fig.~\multiref{#1}}
\newcommand{\figref}[1]{Fig.~\multiref{#1}}
\renewcommand{\eqref}[1]{(\multiref{#1})}

\ifx\href\asklfhas\newcommand{\href}[2]{#2}\fi

\newcommand{\be}{\begin{eqnarray}}
\newcommand{\ee}{\end{eqnarray}}

\makeatletter
\let\old@startsection=\@startsection
\let\oldl@section=\l@section
\renewcommand{\@startsection}[6]{\old@startsection{#1}{#2}{#3}{#4}{#5}{#6\mathversion{bold}}}
\renewcommand{\l@section}[2]{\oldl@section{\mathversion{bold}#1}{#2}}
\makeatother
\usepackage{color}

\begin{document}

\preprint{BONN-TH-2025-20}


\title{Non-Local Symmetries of Planar Feynman Integrals}

\author{Florian Loebbert}
\email{loebbert@uni-bonn.de}
%
\author{Lucas R\"uenaufer}
\email{lucas@rueenaufer.de}
%
\author{Sven F.\ Stawinski}
\email{sstawins@uni-bonn.de}
\affiliation{Bethe Center for Theoretical Physics, Universit\"at Bonn, D-53115, Germany.}

\begin{abstract}
We prove the invariance of scalar Feynman graphs of any planar topology under the Yangian level-one momentum symmetry given certain constraints on the propagator powers. The proof relies on relating this symmetry to a planarized version of the conformal simplices of Bzowski, McFadden and Skenderis. In particular, this proves a momentum-space analogue of the position-space conformal condition on propagator powers. When combined with the latter, the invariance under the level-one momentum implies full Yangian symmetry of the considered graphs. These include all scalar Feynman integrals for which a Yangian symmetry was previously demonstrated at the level of examples, e.g.\ the fishnet or loom graphs, as well as generalizations to graphs with massive propagators.
\end{abstract}

\maketitle

\begin{figure}[t]
\begin{center}
\includegraphicsbox[scale=1]{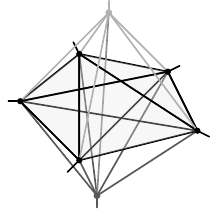}
\quad\tikz[>=latex]{\draw[->,thick,gray] (0,0)--(.5,0);}\quad
\includegraphicsbox[scale=1]{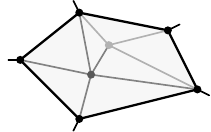}
\end{center}
\caption{Planarizing a momentum-space Feynman graph.}
\label{fig:Planarize}
\end{figure}

Since the pivotal work of 't~Hooft \cite{tHooft:1973alw} it is clear that the planar limit contributes a great deal to our understanding of quantum field theory.  This becomes particularly apparent in the context of the AdS/CFT duality, where integrability emerges in the 't~Hooft limit of a higher dimensional conformal field theory \cite{Beisert:2010jr}. The associated computational tools provide access to QFT data that is out of reach of traditional methods. 
The latter typically rely on the computation of Feynman integrals, whose mathematical structures are currently the subject of great research efforts \cite{Weinzierl:2022eaz,Travaglini:2022uwo}.
These developments are driven from various directions, including precision particle physics, gravitational wave phenomenology or connections to pure mathematics. 
In this letter we study the emergence of integrability when `planarizing' Feynman integrals.  We start with the case of generic (non-planar) simplex integrals  that were shown to be invariant under a momentum-space conformal symmetry~\cite{Bzowski:2019kwd,Bzowski:2020kfw}. The edges of these simplices represent momentum-space propagators, while faces correspond to loop integrations. Reducing this setup via a planarization prescription, cf.\ \Figref{fig:Planarize}, we show that the original momentum-space conformal invariance implies a non-local Yangian level-one momentum symmetry of the resulting dual (or region-momentum) graphs. In the presence of position space loops, this symmetry requires certain constraints on the propagator powers, which follow from our prescription for planarizing simplex graphs via soft limits of external momenta. We note that in
\cite{Kazakov:2023nyu,Levkovich-Maslyuk:2024zdy} these constraints were formulated on the basis of examples. When imposing also dual conformal symmetry via additional constraints on the propagator powers, this implies invariance under the full conformal Yangian, and thus integrability. We extend this setup to include configurations with massive propagators, thus unifying all previous instances where a Yangian symmetry of scalar Feynman integrals was observed \cite{Chicherin:2017cns,Chicherin:2017frs,Loebbert:2019vcj,Loebbert:2020hxk,Kazakov:2023nyu,Loebbert:2024qbw,Levkovich-Maslyuk:2024zdy}. In particular, this includes the fishnet, loom and massive graphs associated to particular QFT models \cite{Gurdogan:2015csr,Kazakov:2018qbr,Loebbert:2020tje,Kazakov:2022dbd,Alfimov:2023vev}
for which Yangian invariance was shown on a case-by-case basis, e.g.\ by using the so-called `lasso method'.
Before proving the announced results, we briefly review some input.

\paragraph{Conformal Invariance of Mesh Integrals.}
The first fact we will need is the conformal invariance of so-called \emph{mesh integrals}~\cite{Bzowski:2020kfw}, which represent a special case of simplex integrals and are defined as 
\begin{align}
	M^{\delta}_n(p_1,\dots ,p_n)=
	&\prod_{1\leq i<j\leq n}C_{ij}\int\frac{\dd^D q_{ij}}{\pi^{D/2}}\frac{1}{q_{ij}^{2(\alpha_{ij}+D/2)}}\nonumber\\
	&\times\prod_{k=1}^n(2\pi)^D\delta^{(D)}\left(p_k+\sum_{i=1}^nq_{ik}\right) \,,
	\label{eq:DeltaMesh}
\end{align}
where the $q_{ij}$ satisfy $q_{ij}=-q_{ji}$. Hence, these are (momentum-space) Feynman integrals corresponding to $n$-gon graphs with an external momentum entering at each vertex and every vertex connected to every other vertex by an internal propagator. 
The power of the propagator which connects vertices $i$ and $j$ is given by $\alpha_{ij}+D/2$. The normalization factors are defined as
\begin{equation}
	C_{ij}=\frac{4^{\alpha_{ij}}}{\Gamma(-\alpha_{ij})}\Gamma\left(\frac{D}{2}+\alpha_{ij} \right) \,,
\end{equation}
with the Gamma-function in the denominator playing a regularizing role below.

We will mostly focus on the reduced mesh integrals $M_n$, which are stripped from the overall momentum conserving delta function and depend only on the external momenta $p_1,\dots ,p_{n-1}$:
\begin{equation}
	M^{\delta}_n(p_1,\dots ,p_n)=(2\pi)^D\delta^{(D)}\left(\sum_{i=1}^np_i\right)M_n(p_1,\dots ,p_{n-1}) \,.
\end{equation}
The crucial information about mesh integrals that we will need in the following, is that they are conformally invariant, i.e., they are annihilated by the (momentum-space) conformal algebra for an appropriate choice of (momentum-space) scaling dimensions $\bar{\Delta}_i$. This invariance can be formulated on the reduced mesh integral, which satisfies the special conformal Ward identity~\cite{Bzowski:2020kfw}
\begin{equation}
	\gen{\bar K}^{\mu}M_n(p_1,\dots ,p_{n-1}) =0 \,,
\end{equation}
with the generator given by
\begin{equation}
\label{eq:momspaceK}
	\gen{\bar K}^{\mu}=\sum_{i=1}^{n-1}\gen{\bar K}_i^{\mu}=\sum_{i=1}^{n-1}\left(p_i^{\mu}\partial_{p_i}^2-2p_i^{\nu}\partial_{p_i,\nu}\partial_{p_i}^{\mu}-2\bar{\Delta}_i\partial_{p_i}^{\mu}\right) \,.
\end{equation}
For this invariance, the (momentum-space) scaling dimensions have to be chosen as  \cite{Bzowski:2020kfw}
\begin{equation}
	\bar{\Delta}_i=D+\sum_{j=1}^n\alpha_{ij},\qquad i=1,\dots ,n-1 \,,
	\label{eq:Deltabar}
\end{equation}
where we defined $\alpha_{ii}=0$ and $\alpha_{ij}=\alpha_{ji}$ for $i>j$. 
Note that when formulating the Ward identity on the full mesh integral, including the momentum conserving delta function, also the $n$th scaling dimension $\bar{\Delta}_n$ enters and is fixed by the dilatation Ward identity, see \emph{App.~A}.
\paragraph{From Momentum to Region Momentum Space.}
The second fact we will need is the connection of the momentum-space special conformal generator to the non-local Yangian level-one momentum generator $\levo{P}^\mu$ in dual momentum coordinates $x_i^\mu$. To this end consider some function $I_n(p_1,\dots ,p_{n-1})$ of $n-1$ linearly independent external momenta, which is annihilated by the special conformal generator for fixed scaling dimensions $\bar{\Delta}_i$: $\gen{\bar K}^{\mu}I_n(p_1,\dots ,p_{n-1})=0$.
Now consider the transformation $p_i=x_i-x_{i+1}$ to the region momentum variables $x_i$ for $i=1,\dots ,n-1$. %
Note that this map requires an ordering of the external legs, and in the following $I_n$ will be identified with a planarized mesh function $M_n$. Interpreting the function $I_n(p_1,\dots ,p_{n-1})$ as a function $I_n(x_1,\dots, x_n)$ of the region momenta and implementing the transformation on the special conformal generator, the special conformal Ward identity turns into the $\levo{P}$ Ward identity \cite{Loebbert:2020hxk}%
\footnote{Note that we discard an additional term proportional to a $p_n$-derivative, which annihilates the delta-stripped integral.}
\begin{equation}
	\levo{P}^{\mu}I_n(x_1,\dots ,x_{n})=0 \,.
\end{equation}
Here the non-local level-one momentum operator is defined as
\begin{equation}
	\levo{P}^{\mu}=\sfrac{i}{2}\sum_{1\leq j<k\leq n}\left[\gen{P}_j^{\mu}\gen{D}_k+\gen{P}_{j\nu}\gen{L}_k^{\mu\nu}-(j\leftrightarrow k)\right]+\sum_{j=1}^n s_j \gen{P}_j^{\mu} \,,
	\label{eq:DefPhat}
\end{equation}
in terms of the conformal momentum, Lorentz and dilatation generators, respectively:
\begin{align}
	\gen{P}_j^{\mu}&=-i\partial_{x_j}^{\mu} \,, 
	\qquad
	\gen{L}_j^{\mu\nu}=i(x_j^{\mu}\partial_{x_j}^{\nu}-x_j^{\nu}\partial_{x_j}^{\mu}),
	 \\
	 \gen{D}_j&=-i(x_{j\mu}\partial_{x_j}^{\mu}+\Delta_j)
	 \nonumber\,.
\end{align}
Together with the $x$-space conformal Lie algebra generators, the operator $\levo{P}^\mu$ generates the infinite dimensional Yangian algebra, which underlies the integrability of rational integrable models (see e.g.\ \cite{Loebbert:2016cdm} for a review). $\levo{P}^\mu$ depends on the scaling dimensions $\Delta_i$ and the so-called evaluation parameters~$s_i$. These are related to the momentum space scaling dimensions $\Bar{\Delta}_i$ through \cite{Loebbert:2020hxk}:
\begin{equation}
	\bar{\Delta}_i=\frac{1}{2}(\Delta_i+\Delta_{i+1}+2s_i-2s_{i+1}) , \qquad i=1,\dots ,n-1 \,.
	\label{eq:DeltaBarEval}
\end{equation}
Since we are not necessarily considering functions invariant under the position space conformal algebra, the scaling dimensions $\Delta_j$ are merely some parameters which could be absorbed into the evaluation parameters. In the following we choose $\Delta_j$ to equal the power of the external propagator connected to the point $x_j$.

\section{Proof of $\levo{P}$ Invariance}
We will now prove that all planar, massless (position-space) Feynman graphs with generic external kinematics are annihilated by the $\levo{P}$ operator, as long as specific constraints on the propagator powers hold. 
\subsection{Tree Graphs from Propagator-Power Limits}

\begin{figure}[t]
\begin{center}
\begin{tabular}{ccc}
Mesh&Tree&Loop
\\
\includegraphicsbox[scale=.7]{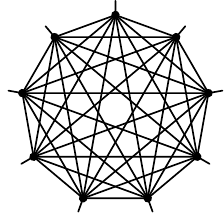}
&
\includegraphicsbox[scale=.7]{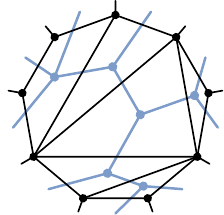}
&
\includegraphicsbox[scale=.7]{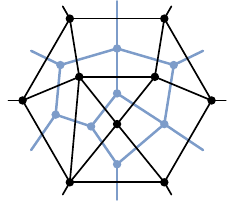}
\\
$M_n$
&
$\lim\limits_{\alpha_{ij}\to 0}M_n$
&
$\lim\limits_{\substack{\alpha_{ij}\to 0\\p_{1,2,3}\to 0}} M_n$
\end{tabular}
\end{center}
\caption{Left: A generic nine-point mesh integral in momentum space. Middle: Removing certain propagators from the mesh, one obtains a nine-point graph, which is dual to a position-space tree (blue). Right: A six-point integral dual to a position-space loop graph (blue), and obtained by removing certain propagators from the generic mesh and taking three external momenta soft; the latter three vertices are moved to the inside of the black graph after the soft limit.}
\label{fig:meshred}
\end{figure}
First consider a generic planar tree-level graph with position-space representation
\begin{equation}
	I_n(x_1,\dots ,x_n)=\prod_{k=n+1}^{n+L} \int\frac{\dd^D x_k}{\pi^{D/2}}\prod_{e \in E}\frac{1}{x_{e}^{2a_{e}}} \,,
\end{equation}
for generic propagator powers $a_{e}$.
Here $E$ is a set of edges and $x_e^2=(x_i-x_j)^2$ denotes the squared distance between the end points $i,j$ of edge $e$. 
Under the transformation $p_i=x_i-x_{i+1}$, we can reinterpret this integral as a function of $n-1$ momenta $I_n(p_1,\dots ,p_{n-1})$, with the corresponding graph being the dual graph of the original graph. 
Since the original graph was a tree graph, this dual graph is outerplanar, i.e.\ all of its vertices lie on the boundary. 
This is true by definition of graph duality. 
Note that all vertices are connected to the incoming momenta $p_1,\dots p_{n-1},p_n=-\sum_{i=1}^{n-1}p_i$.

The fact that the momentum space graph, which is dual to our position space tree, is outerplanar, guarantees that the corresponding momentum-space Feynman integral corresponds to a particular mesh integral introduced in the previous section with some propagators removed, see \Figref{fig:meshred}. The latter is easily implemented on the mesh integral by sending the corresponding $\alpha$ parameter(s) to zero: $\alpha_{ij}\to0$ for $(ij)\in A$, where $A$ is the set of propagators to be removed. The reason is the following identity
\begin{equation}
	\lim_{\alpha\rightarrow 0}\frac{1}{\Gamma(-\alpha)}\frac{1}{q^{2(D/2+\alpha)}}=\frac{\pi^{D/2}}{\Gamma\left(\frac{D}{2}\right)}\delta^{(D)}(q) \,,
\end{equation}
which, under the $q_{ij}$ integral in \eqref{eq:DeltaMesh}, implies that the limit $\alpha_{ij}\rightarrow 0$ removes the corresponding propagator from the mesh integral. Hence we can view the (position-space tree) integral  $I_n$ as a (reduced) mesh integral 
\begin{equation}
	I_n(p_1,\dots ,p_{n-1}) =\lim_{\substack{\alpha_{ij}\rightarrow 0 \\ (ij)\in A}} M_n(p_1,\dots,p_{n-1}) \,.
\end{equation}
The nonzero $\alpha_{ij}$ parameters are determined by the propagator powers of the Feynman integral via $\alpha_{ij}=a_k-D/2$ for some appropriate $k$. 

Now the proof of $\levo{P}$ symmetry of the original tree Feynman integral $I(x_1,\dots ,x_n)$ is quite straightforward. After reading off the $\alpha_{ij}$ from the graph, we fix the (momentum-space) scaling dimensions, which relate to the evaluation parameters via
\eqref{eq:DeltaBarEval}.
Combining the results reviewed in the previous section, we then have
\begin{align}
	0&=\gen{\bar{K}}^{\mu} \lim_{\substack{\alpha_{ij}\rightarrow 0 \\ (ij)\in A}} M_n(p_1,\dots,p_{n-1}) \nonumber \\
	&=\gen{\bar{K}}^{\mu} I_n(p_1,\dots ,p_{n-1}) \\
	&=\levo{P}^{\mu}I_n(x_1,\dots ,x_n) \nonumber \,.
	\end{align}
Note that in the above steps we have implicitly commuted the special conformal generator with the $\alpha_{ij}\rightarrow 0$ limit; the limit just fixes the scaling dimensions in the generator and hence does not affect the invariance. 
\subsection{Loop Graphs from Soft Limits}

Let us now consider general planar position-space graphs, including (position-space) loops.  A loop in position space represents an internal face, which under graph duality turns into an internal vertex in momentum space. A priori internal vertices do not exist within a mesh integral, and we will incorporate this  feature into the (generically non-planar) mesh integrals by taking soft limits. Explicitly, if we take an external momentum $p_k$ of some (reduced) mesh integral to zero, this means that no external momentum flows into this vertex, and hence this vertex becomes an internal vertex; to stress this graphically, we move such vertices to the inside of the polygon, see \Figref{fig:meshred} for an illustration.
We emphasize that here the notion of loops includes closed faces obtained by external coincidence limits, cf.\ \emph{App.~B}.

We start from some planar position-space Feynman integral $I_n^{(\ell)}(x_1,\dots ,x_n)$ with $\ell$ position-space loops.  We rewrite this integral in terms of independent external momenta by interpreting the $x_i$ as region momenta and viewing the resulting integral $I_n^{(\ell)}(p_1,\dots ,p_{n-1})$ as the soft limit of a (reduced) mesh integral:
\begin{equation}
	I_n^{(\ell)}(p_1,\dots ,p_{n-1})=\hspace{-4mm}\lim_{\substack{p_i\rightarrow 0 \\ i\in\{n,\dots ,N-1\}}}\lim_{\substack{\alpha_{ij}\rightarrow 0 \\ (ij)\in A}}M_{N}(p_1,\dots ,p_{N-1}) \,,
\end{equation}
where we abbreviate 
\begin{equation}
	N\equiv N(n,\ell):=n+\ell \,,
\end{equation}
for fixed $n,\ell$. Recall that $A$ is the set of propagators removed from the mesh.

Now, as for the tree graphs, we would like to use the fact that the mesh is annihilated by the special conformal generator to show the $\levo{P}$ symmetry of the original Feynman integral. However, we only know the symmetry before taking the soft limit, but want to use it afterwards. 
We can commute the limit with the generator as long as the mesh integral and its first and second derivatives are finite in this limit (this is certainly true for some choice of propagator powers, the symmetry can then be extended to other choices via analytic continuation) and we choose the corresponding momentum-space scaling dimensions to be zero. These conditions are directly inferred from the explicit form of the special conformal generator density, see \eqref{eq:momspaceK}. Namely, for a well-behaved function $f(p_i)$ we have
\begin{equation}
	\lim_{p_i\rightarrow 0}\gen{\bar{K}}_i^{\mu}f(p_i)=-2\bar{\Delta}_i\lim_{p_i\rightarrow 0}\partial_{p_i}^{\mu}f(p_i) \,.
\end{equation}

Note, however, that in general, setting $\bar{\Delta}_i=0$ is not possible, since the $\bar{\Delta}_i$ are already fixed in terms of the propagator powers through the relations \eqref{eq:Deltabar}.
Hence, we find that the symmetry holds if the propagator powers satisfy the following linear relations:
\begin{equation}
	D+\sum_{j=1}^{N}\alpha_{ij} =0 ,\qquad i=n,\dots N-1 \,.
	\label{eq:PhatConstraints}
\end{equation}

Let us for now assume that these relations are satisfied, i.e.\ that $\bar{\Delta}_i=0$ for $i=n,\dots, N-1$. As argued above, this allows us to commute limits $p_i\rightarrow 0$ with the action of the special conformal generators for $i=n,\dots, N-1$. Hence we can show the $\levo{P}$ invariance of the Feynman integral $I^{(\ell)}(x_1,\dots ,x_n)$ following a similar argument as before: 
\begin{align}
	0&= \lim_{p_n\rightarrow 0}\dots \lim_{p_{N-1}\rightarrow 0} \lim_{\substack{\alpha_{ij}\rightarrow 0 \\ (ij)\in A}}\gen{\bar{K}}^{\mu}M_N(p_1,\dots ,p_{N-1}) \nonumber\\
	&=\gen{\bar{K}}^{\mu} \lim_{p_n\rightarrow 0}\dots \lim_{p_{N-1}\rightarrow 0} \lim_{\substack{\alpha_{ij}\rightarrow 0 \\ (ij)\in A}}M_N(p_1,\dots ,p_{N-1}) \nonumber \\
	&=\gen{\bar{K}}^{\mu}I_n^{(\ell)}(p_1,\dots ,p_{n-1})  \nonumber \\
	&=\levo{P}^{\mu}I_n^{(\ell)}(x_1,\dots ,x_{n}).
\end{align}
Note that in the first line the special conformal generator acts on the independent momenta $p_1,\dots ,p_{N-1}$ while in the second and third lines the generator acts on the independent momenta $p_1,\dots, p_{n-1}$. Also recall that the evaluation parameters are fixed in terms of the momentum-space scaling dimensions via  \eqref{eq:DeltaBarEval}.
\subsection{Propagator Powers and Evaluation Parameters}
Let us take a closer look at the constraints \eqref{eq:PhatConstraints} for the propagator powers, which are required for invariance under the level-one momentum generator.
We note that there is one constraint for every position-space loop. Consider now a fixed position-space loop, which takes the form of a $k$-gon and corresponds to the internal vertex $i$ in momentum space. The sum of $\alpha$ parameters entering the constraint takes the form of a sum over all propagators connected to the vertex $i$ in momentum space, which corresponds to all edges of the $k$-gon in position space. Let us denote the powers of these propagators by $a_1,\dots, a_k$. Then the above proof implies
\begin{equation}
	\includegraphicsbox[scale=1]{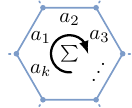}=\sum_{m=1}^k a_m=\frac{(k-2)D}{2} \,,
	\label{eq:momconfconstraints}
\end{equation}
for each position-space loop in the graph, which are precisely the constraints that were found for various examples in \cite{Levkovich-Maslyuk:2024zdy}.

Using a similar argument, we show in \emph{App.~C} that the evaluation parameters are computed recursively as 
\begin{equation}
	s_{i+1}=s_i-\frac{1}{2}(a_i+a_{i+1})-\sum_{j}\left(b_j-\frac{D}{2}\right)\,,
	\label{eq:evalParams}
\end{equation}
where we can start the recursion with an arbitrary choice of $\Eval_1$, since the evaluation parameters are only defined up to an overall shift by the total momentum. Here the $b_j$ are the powers of internal propagators that one has to traverse when following the boundary of the position space graph from $x_i$ to $x_{i+1}$ and $a_i,a_{i+1}$ are the powers of the external propagators connected to $x_i,x_{i+1}$, respectively. 
This formula has been motivated in \cite{Chicherin:2017cns,Loebbert:2020hxk,Kazakov:2023nyu} and proven for generical tree-graphs in \cite{Loebbert:2024qbw}. Our proof here extends to graphs with position-space loops and hence proves the relation in full generality.
\subsection{Massive Meshes and Level-One Momentum}

\begin{figure}[t]
\begin{center}
\begin{tabular}{ccc}
\includegraphicsbox[scale=.65]{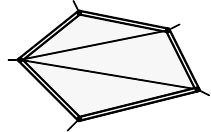}\;\tikz[>=latex]{\draw[->,thick,gray] (0,0)--(.4,0);}\;
&
\includegraphicsbox[scale=.65]{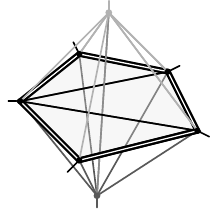}\;\tikz[>=latex]{\draw[->,thick,gray] (0,0)--(.4,0);}\;
&
\includegraphicsbox[scale=.65]{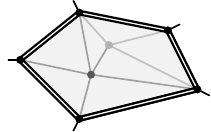}
\\
Step 1\phantom{aaa}&Step 2\phantom{aaa}& Step 3
\end{tabular}
\end{center}
\caption{Proof of the massive level-one momentum symmetry in terms of momentum-space graphs. We start with a planar mesh with massive boundary, whose dual (tree) graph is invariant under $\mathbb{\levo{P}}^\mu$ alias momentum-space $\mathbb{\bar K}^\mu$ \cite{Loebbert:2024qbw}. Next we use the induction of \cite{Bzowski:2020kfw}, to extend the $\mathbb{\bar K}^\mu$ invariance to non-planar massive cycle mesh integrals. Finally, the latter are planarized via soft limits, which preserve the $\mathbb{\levo{P}}^\mu$ symmetry and lead to (position-space loop) integrals with massive boundary.}
\label{fig:massivecycle}
\end{figure}

In \cite{Loebbert:2020hxk,Loebbert:2020glj} it was argued  that the $\levo{P}^{\mu}$ symmetry can be extended to graphs with massive external propagators of the form $1/(x_{e}^2+m_{e}^2)$. This statement was proved in \cite{Loebbert:2024qbw} for generic position-space \emph{tree graphs}, which will serve us as the base case for an induction. The massive generalization $\widehat{\mathbb{P}}^{\mu}$ of $\levo{P}^{\mu}$ takes again the form \eqref{eq:DefPhat}, with the dilatation generator modified by a mass term, cf.\ \cite{Alday:2009zm}:
\begin{equation}
	\mathbb{D}_j=-i(x_{j\mu}\partial_{x_j}^{\mu}+\Delta_j+m_j\partial_{m_j}) \,.
\end{equation}
Here $m_j$ is the mass of the propagator connected to $x_j$. Translating $\widehat{\mathbb{P}}^{\mu}$ to dual momentum space yields a massive generalization of the special conformal generator~\cite{Loebbert:2020hxk}
\begin{equation}
	\mathbb{\bar{K}}_j^{\mu}=\gen{\bar K}^{\mu}-(m_j\partial_{m_j}+m_{j+1}\partial_{m_{j+1}})\partial_{p_j}^{\mu} \,.
\end{equation}
The correspondence between position-space (tree) graphs and momentum-space meshes discussed above then immediately implies that mesh integrals with massive boundary propagators are invariant under the massive generalization of the conformal algebra --- at least as long as they are reduced to planar graphs by appropriate choice of $\alpha$-parameters, see step~1 in \Figref{fig:massivecycle}.

Next we extend this invariance under the massive special conformal generator to a certain class of \emph{massive cycle mesh} integrals. 
The graphs in this class take the form of an $n$-gon with massive boundary propagators embedded into a generic $N$-vertex mesh integral,  cf.\ step~2 in \Figref{fig:massivecycle}. Vertices that lie on the $n$-gon are connected among each other in a planar fashion only, with the $\alpha_{ij}$
 fixed to guarantee this. There is however no restriction on the edges connected to the vertices $n+1,\dots ,N$.

We can reach such massive cycle mesh integrals recursively by starting with an $n$-gon mesh with massive boundary and adding vertices connected to all other vertices via massless propagators. Explicitly we have  \cite{Bzowski:2020kfw}
\begin{align}
	&\mathbb{M}^{\delta}_n(p_1\dots,p_n)=\prod_{i=1}^{n-1} C_{in}\int\frac{\dd^D q_i}{\pi^{D/2}}\frac{1}{\prod_{j=1}^{n-1}q_j^{2(\alpha_{jn}+D/2)}} \nonumber \\
	& \times \mathbb{M}^{\delta}_{n-1}(\hat p_1, \dots,\hat p_{n-1})(2\pi)^D\delta\left( p_n+\sum_{j=1}^{n-1}q_j\right)  \,,
\end{align}
where $\hat{p}_i=p_i-q_i$ and both, $\mathbb{M}_n^{\delta}$ and $\mathbb{M}^{\delta}_{n-1}$, are massive cycle mesh integrals. It was shown in \cite{Bzowski:2020kfw} that the massless special conformal generator can be pulled through the integral to act only on the smaller mesh integral. Since the massive special conformal generator is a sum of the massless generator and mass derivatives, the massive generator can trivially be pulled through the propagators as well. Hence, in precisely the same way as in \cite{Bzowski:2020kfw}, the massive special conformal invariance of $\mathbb{M}_n$ is reduced to the one of $\mathbb{M}_{n-1}$. Every massive cycle mesh can now be reached in this way from a simple $n$-gon with massive boundary, for which this symmetry was already established in step~1. Hence the recursive argument allows us to extend the symmetry to all massive cycle mesh integrals inductively.

In the final step~3 we now employ the same argument as before and interpret the soft limit\footnote{Note that the commutation of the generator with the soft limit is not affected since the generators acting on the momenta which we take soft do not involve the masses.} of a massive cycle mesh integral as a momentum-space Feynman integral dual to an arbitrary planar position-space integral with the massive special conformal generator transforming into the massive $\widehat{\mathbb{P}}^{\mu}$ generator. 
This proves that all planar position-space Feynman integrals with massive external legs feature this massive symmetry. 

\section{Outlook}

An important question following the above observations concerns the constraining power of the $\levo{P}$ symmetry. For three external points it is well known that position-space conformal symmetry implies that any correlator reduces to a product of free propagators (a triangle). Similarly, $\levo{P}$ symmetry (without imposing $x$-space conformal symmetry) implies that any three-point integral obeying the constraints \eqref{eq:momconfconstraints}, reduces to a linear combination of Appell $F_4$ hypergeometric functions, which constitute the generic star integral, cf.\ \cite{Boos:1987bg,Coriano:2013jba,Bzowski:2013sza,Coriano:2019sth,Loebbert:2019vcj,Loebbert:2020hxk}:
\begin{equation}
\includegraphicsbox[scale=.8]{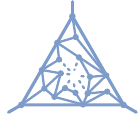}
\simeq
\includegraphicsbox[scale=.8]{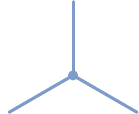}
\end{equation}
This can be attributed to the fact that the $\levo{P}$ constraints \eqref{eq:momconfconstraints} allow us to fully express the powers of propagators in the bulk of the Feynman diagram in terms of boundary propagators. Put differently, the Yangian level-one momentum only `sees' the boundary of the graph, cf.\ \emph{App.~C}. 
Notably, position-space \emph{tree} graphs are `pure boundary' and thus $\levo{P}$-invariant for \emph{any} choice of propagator powers. In particular, this includes generic track integrals, e.g.\  
\begin{equation}
\includegraphics[scale=1]{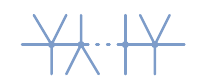}
\end{equation}
which generalize the train track and triangle track graphs of \cite{Bourjaily:2018ycu,Vergu:2020uur,Duhr:2022pch,Cao:2023tpx,McLeod:2023doa,Duhr:2024hjf} to the non-conformal case.
It will be interesting to further explore the role of the $\levo{P}$ differential equations for defining families of integrals at higher numbers of points, cf.\ \cite{Duhr:2024hjf, Ferrando:2025duw}.

It should be stressed that position space trees are even invariant under more elementary non-local operators  \cite{Loebbert:2024qbw}, whose sum constitutes the full generator $\levo{P}$. It should be clarified how these sub-symmetries relate to the simplex approach and which of them persist for the loop graphs considered in this letter.


Another important direction concerns the super-symmetrization of the approach followed here: Can the simplex construction of momentum-space conformal invariants be super-symmetrized and connected to the integrability of planar $\superN=4$ Super Yang--Mills Theory? Including the brick wall fishnet integrals with fermionic propagators  into the present discussion could be a first step towards this goal, cf.\ \cite{Chicherin:2017frs,Pittelli:2019ceq,Kade:2023xet,Kade:2024ucz,Kade:2024lkc}.

Finally, we ask whether the integrability of Feynman integrals is introduced via the process of planarization, or whether a non-planar version of integrability is present in the original simplex graphs. A natural starting point to approach this question is Zamolodchikov's tetrahedron generalization of the Yang--Baxter equation~\cite{zamolodchikov1980tetrahedra}.

\medskip

\begin{acknowledgments}
\noindent\emph{Acknowledgements:}  
We are grateful to Claude Duhr and Gwena\"el Ferrando for helpful discussions and we thank Gwena\"el Ferrando for useful comments on the manuscript. The work of SFS is funded by the European Union (ERC Consolidator Grant LoCoMotive 101043686).
Views and opinions expressed are however those of the author(s) only and do not necessarily reflect
those of the European Union or the European Research Council. Neither the European Union nor the
granting authority can be held responsible for them. The work of FL is funded by the Deutsche Forschungsgemeinschaft (DFG, German Research Foundation) -- Projektnummer  508889767.
\end{acknowledgments}
\bibliography{PhatFromMomSpace}

\section{Appendix}

\paragraph{Appendix A: Conformal Symmetry - Full vs Reduced Meshes.}
\label{sec:specialconfcomm}
In this appendix we give some details on the relation between the conformal invariance of full mesh integrals and their reduced (delta-stripped) versions, see also \cite{Bzowski:2020kfw}. To relate the action of the special conformal generator on the full and on the reduced mesh integral, we need to consider the commutator of the special conformal generator (acting on all momenta $p_1,\dots ,p_n$) with the momentum conserving delta function. By using some distributional identities we can compute 
\begin{equation}
	[\gen{\bar{K}}^{\mu},\delta(P)]=2\partial_{\nu}\delta(P) \gen{\bar L}^{\mu\nu}-2\partial^{\mu}\delta(P)\left( \gen{\bar D}-D \right),
	\label{eq:deltacommutator}
\end{equation}
where we use $P=\sum_{i=1}^n p_i$ and the derivatives of the delta-functions are taken with respect to $P^{\mu}$.  Moreover, we employ the momentum-space generators
\begin{align}
	\gen{\bar{L}}^{\mu\nu}&=\sum_{j=1}^{n}(p_j^{\mu}\partial_{p_j}^{\nu}-p_j^{\nu}\partial_{p_j}^{\mu}) \,, \\
	\gen{\bar{D}}&=\sum_{j=1}^{n}(p_j^{\mu}\partial_{p_j}^{\mu}+\bar{\Delta}_j) \,. \nonumber
\end{align}
Since the reduced mesh integral is Lorentz invariant, we can relate the special conformal Ward identity of the full mesh integral to the one of the reduced mesh by choosing $\bar{\Delta}_n$ such that the second term in \eqref{eq:deltacommutator} is zero. Simply counting the dimension of the mesh integral shows that the choice 
\begin{equation}
	\bar{\Delta}_n=D+\sum_{i=1}^n\alpha_{in} \,,
\end{equation}
guarantees this. 
\paragraph{Appendix B: Coincidence Limits.}
In the main text we noted that our notion of loops includes the closed faces introduced by coincidence limits of external points. Hence our results in particular imply the $\levo{P}$ symmetry for graphs with coincidence limits as long as the constraints \eqref{eq:momconfconstraints} are satisfied. Note however that these constraints often allow the simplification of the graphs through application of the star-triangle identity
\begin{equation}
	\int\frac{\dd^D x_0}{\pi^{\frac{D}{2}}}\frac{1}{x_{01}^{2a}x_{02}^{2b}x_{03}^{2c}}=\frac{X_{abc}}{x_{12}^{2c'}x_{23}^{2a'}x_{13}^{2b'}} \,,
\end{equation}
with
\begin{equation}
	X_{abc}=\frac{\Gamma\left(a'\right)\Gamma\left(b'\right)\Gamma\left(c'\right)}{\Gamma(a)\Gamma(b) \Gamma(c)} \,,
\end{equation}
and $a'=D/2-a$, etc. This relation holds only if $a+b+c=D$, which implies $a'+b'+c'=D/2$.

\begin{figure}[t]
\begin{center}
\includegraphicsbox[scale=1]{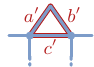}
$\simeq$
\includegraphicsbox[scale=1]{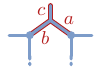}
\qquad\qquad
\includegraphicsbox[scale=1]{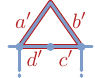}
\end{center}
\caption{Left: Coincidence limit that can be removed for the $\levo{P}$ constraint $a'+b'+c'=D/2$ by virtue of the star-triangle relation. Right: Graph with a coincidence limit that cannot be removed using the star-triangle identity but features a $\levo{P}$ symmetry for $a'+b'+c'+d'=D$.}
\label{fig:coincidence}
\end{figure}

Consider for example a graph with two points connected by one internal propagator and connected to the same external point as in \Figref{fig:coincidence} (left). Our results show that this graph has the $\levo{P}$ symmetry as long as the propagator powers along the triangle making up the coincidence limit sum up to $D/2$ (and possibly other conditions from the rest of the graph are satisfied). This is precisely the condition which allows us to use the star-triangle identity to get rid of the coincidence limit.

There are however also examples of non-trivial coincidence limits for which we have proven the $\levo{P}$ symmetry. Consider for example a graph where two vertices are connected to the same external point but are separated by two internal propagators, see \Figref{fig:coincidence} (right). Then we need to impose that the four propagator powers making up the quadrilateral sum up to $D$. Hence we can not get rid of the coincidence limit (in an obvious way).

\paragraph{Appendix C: Proof of Formula for Evaluation Parameters.}
\label{app:evalParamProof}
In this appendix we will prove the relation for the evaluation parameters \eqref{eq:evalParams}, stated in the main text. The equation connecting the evaluation parameters to the momentum-space scaling dimensions \eqref{eq:DeltaBarEval} (with $\Delta_i=a_i$) can be rewritten recursively as
\begin{equation}
	s_{i+1}=s_i+\frac{1}{2}(a_i+a_{i+1})-D-\sum_{j=1}^{N} \alpha_{ij} \,.
	\label{eq:EvalRec}
\end{equation}
To find a useful formula, we hence need to understand the sum over $\alpha$ parameters.

\begin{figure}[t]
\begin{center}
\includegraphicsbox[scale=1]{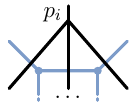}
\quad$\leftrightarrow$\quad
\includegraphicsbox[scale=1]{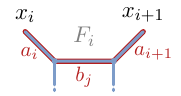}
\end{center}
\caption{Illustration on the prescription for reading off the evaluation parameters by going along the boundary of the graph in $x$-space (blue) or its dual (black).}
\label{fig:evalparams}
\end{figure}

We are summing over all edges of the momentum-space graph which end at a particular vertex $i$, at which the external momentum $p_i$ is inflowing. In region momentum space, this vertex becomes a face $F_i$, namely the external face bounded by the propagator line between points $x_i$ and $x_{i+1}$. The dual of the set of edges connected to point $p_i$, is the set of edges of the region-momentum graph which border this face. Hence the edges we are summing over can be described by starting with point $x_i$ and traversing the boundary of the graph until reaching $x_{i+1}$. The sum over $\alpha$ parameters then splits up into the two external propagators connected to $x_i$ and $x_{i+1}$ with propagator powers $a_i,a_{i+1}$, respectively, and all internal propagators that are traversed when following the boundary of the graph. We denote the corresponding internal propagator powers by $b_j$, see \figref{fig:evalparams} for an illustration. The sum over $\alpha$ parameters then yields (recall that edges absent from the diagram have $\alpha_{ij}=0$) 
\begin{equation}
	\sum_{j=1}^{N}\alpha_{ij}=\left(a_i-\frac{D}{2}\right)+\left(a_{i+1}-\frac{D}{2}\right)+\sum_{j}\left(b_j-\frac{D}{2}\right) \,.
\end{equation}
Plugging this into the above formula \eqref{eq:EvalRec} yields the recursive relation for the evaluation parameters \eqref{eq:evalParams}, as claimed.

\end{document}